\documentclass[12pt,preprint]{aastex}

\begin{document}

\title {A Dusty Disk Around GD 362, a  White Dwarf With a Uniquely High Photospheric Metal Abundance}

\author{E. E. Becklin\footnote{Department of Physics and Astronomy and Center for Astrobiology, University of California, Los Angeles CA 90095-1562; becklin, jura, ben@astro.ucla.edu}, J. Farihi\footnote{Gemini Observatory, 670 North A'ohoku Place, Hilo HI 96720; jfarihi, song@gemini.edu}\,\,, M. Jura$^{1}$, Inseok Song$^{2}$, A. J. Weinberger\footnote{Department of Terrestrial Magnetism, Carnegie Institution of Washington, 5241 Broad Branch Road, NW, Washington DC 20015; alycia@dtm.ciw.edu}\,\,, B. Zuckerman$^{1}$}

\begin{abstract}
 Eighteen years after an infrared excess was discovered associated with the white dwarf G29-38, we report  
ground-based  measurements ($JHK_{s}KL'N'$) with  mJy-level sensitivity   of  GD 362 that show  it to be a second  single white dwarf with
an infrared excess.    As a first approximation, the  excess around GD 362, which amounts to ${\sim}$3\% of the total stellar luminosity, can be explained by emission from
a passive, flat, opaque dust disk that lies within the Roche radius of the white dwarf.  The dust may have been produced by the tidal disruption of
a large parent body such as an asteroid.   Accretion from this circumstellar disk could account for the remarkably high abundance of  metals in the star's
photosphere.

\end{abstract}

\section{INTRODUCTION}

It is likely that many planetary systems survive a  star's evolution as a red giant and persist as the  star becomes a white dwarf. Both infrared and optical studies of these systems can constrain the incidence and dynamical evolution of extra-solar comets, asteroids and/or planets.  Here we report  excess  near and mid-infrared emission around the white dwarf GD 362 that
might be explained as the result of the tidal disruption of a parent body such as an  asteroid.     

 Because their initial complement of heavy elements gravitationally settles beneath the photosphere, most cool ($T_{eff}$$<$ 20,000 K) white dwarfs have atmospheres that are expected to be either essentially pure
hydrogen or pure helium.  However,
${\sim}$25\% of cool DA (hydrogen-rich) white dwarfs display at least some  measurable calcium
 in their photospheres (Zuckerman et al. 2003).   Since  the dwell times for atmospheric calcium in 0.6 M$_{\odot}$ and 1.0 M$_{\odot}$ hydrogen-rich white dwarfs with $T_{eff}$ = 10000 K are only ${\sim}$600 years and ${\sim}$40 years, respectively,
 (Paquette et al. 1986), it is likely that these DAZ stars currently are  accreting. 

The  source of the accreting material onto single white dwarfs is not known. One possible source is interstellar
matter (Dupuis et al. 1993). However,  interstellar calcium is largely contained within grains and thus the amount that is  accreted may be small (see Alcock and Illarionov 1980), especially from the low density interstellar medium of the Local Bubble. Furthermore, in the case of Bondi-Hoyle accretion, a correlation would be expected  between photospheric metals and  white dwarf kinematics, yet  none has been found (Zuckerman et al. 2003).   An alternate possibility is that white dwarfs accrete  circumstellar instead of interstellar matter (Aannestad et al. 1993).  One such scenario is that comets directly impact the
photosphere of the white dwarf (Alcock et al. 1986). 

Because their outer convective envelopes are  thin,
metal accretion rates as low as 10$^{6}$ g s$^{-1}$ can account for the abundances of heavy elements in some white dwarfs (see Paquette et al. 1986 and Zuckerman et al. 2003).  The zodiacal light in the Solar System is explained by dust production  from the erosion of comets and asteroids with a rate of 3 ${\times}$ 10$^{6}$ g s$^{-1}$ (Fixsen \& Dwek 2002) and  analogous circumstellar dust debris produced by the destruction of  parent bodies is
common around main sequence stars  (Zuckerman 2001).  It is plausible that  grinding of parent bodies into dust  may 
be on-going around white dwarfs. Because these stars have both low luminosities and low wind-outflow rates,
any orbiting, ground-up dust ultimately can accrete onto its host star and pollute the photosphere.       

With [Ca]/[H] = 1.2 ${\times}$ 10$^{-7}$, G29-38 has the second greatest atmospheric calcium  abundance in the survey of ${\sim}$100 white dwarfs by Zuckerman et al. (2003), and, to date, despite a large amount of effort it has been the only  single white
dwarf with a known  infrared excess (Zuckerman \& Becklin 1987, Farihi et al. 2005). Even observations with the ${\it Infrared\; Space\; Observatory}$  failed to discover any more
 infrared excesses around white dwarfs (Chary et al. 1999).    A plausible
explanation for the excess around G29-38  is that an asteroid strayed within
the tidal radius of the star, broke apart and a cascade of self-collisions  formed a dust disk  that is now accreting onto the star (Jura 2003).    
Recently, Gianninas et al. (2004) found that the  massive white dwarf GD 362   exhibits [Ca]/[H]
= 6 ${\times}$ 10$^{-6}$ which, remarkably,  is even higher than the Solar abundance of 2 ${\times}$ 10$^{-6}$. GD 362 has the highest known photospheric calcium abundance of any white dwarf with $T_{eff}$ $<$ 25000 K (Zuckerman et al. 2003, Koester et al. 2005).   Because of the great advances in infrared instrumention at the Gemini North telescope, it is now possible to achieve ${\sim}$ mJy-level sensitivity at mid-infrared wavelengths. Here we report ground-based detections of GD 362 that show it possesses an infrared  excess that is probably produced by circumstellar dust.  
Our result strengthens the argument that photospheric metals in
white dwarfs result from the accretion of circumstellar matter.

\section{OBSERVATIONS} 
 
We have obtained ground-based observations of GD 362 in three separate observing runs.

On 2005 May 19 and 21 (UT), we observed with an $N'$ (11.3 $\mu$m) filter
in the MICHELLE (Glasse et al. 1997) instrument at the Gemini North
telescope.  Data were obtained in a beam-switching chop-nod mode with a
secondary mirror chop throw of 15${\arcsec}$ at 4.2 Hz.  The total usable on-source
integration time was 2587 s.   The total flux uncertainty was
calculated by summing in quadrature the  measurement (0.24 mJy) and 
calibration (0.1 mJy) uncertainties.   The
calibration uncertainty was derived from the  scatter in the 
five  measurements of the standard star HD 158899 combined with an estimate of the noise from structure due to imperfect background substraction.  Because both science and
standard stars were observed within a narrow range of airmass, 1.0
to 1.3, we did not correct for airmass.  The $N'$  image of GD 362 is shown in Figure 1 and the flux is listed  in Table 1.

Images of GD 362 were taken at $H$ and $K_{s}$ bands with the PANIC instrument
(Martini et al. 2004) on the Magellan I telescope  2005 June 17 (UT). The night was partly
cloudy, the  seeing was a rather poor $\sim$0{\farcs}9. At $K_{s}$ and $H$, a total of
550s and 225s of integration time was collected, respectively.  Data reduction proceeded in the
usual way of combining the dithered observations to make source-free
sky frames, linearizing the data, sky-subtracting each image,
flat-fielding with a sky flat, correcting bad pixels by interpolation,
distortion correcting the images, and shifting and adding the dither
positions together.    Aperture photometry was performed on the white dwarf and on two 
bright stars in the field of view (2MASS 17313587+3705357  and 2MASS
17313555+3704541) with S/N$>$10 detections at $H$ and $K_{s}$ in the
2MASS catalog.  The resulting magnitudes of GD 362 are listed in Table 1 where the stated uncertainties combine the statistical photometric
uncertainty with the calibration uncertainty.

On 2005 June 25 (UT) $JHKL'$ images of GD 362 
were obtained with the Gemini North  telescope  using the  Near-Infrared
Imager, NIRI (Hodapp et al. 2003).   At all wavelengths, images were takin in a four-point dither pattern for total integration times of 80, 100, 120 and 1880 s at $J$, $H$, $K$ and $L'$, respectively. Reduction of the $JHK$ data included sky subtracting
each raw frame by the median of all four images, flat
fielding, image registration, and finally averaging.
Aperture photometry was performed on both GD 362 and
the standard FS 146 (Hawarden et al. 2001)  yielding S/N $>100$
in all bandpasses and  $\sim3$\% uncertainties.  The $L'$ image reductions proceeded similarly to the $JHK$ reductions for the 
standard FS 147 (Leggett et al. 2003).  However,  the $L'$ image reductions for GD 362 were  more complicated since the target  was not visible in any single
frame.  After shifting by the telecope coordinates and coadding, GD 362 appeared in the final image with a S/N ${\sim}3-5$.  The
dominant source of uncertainty is structure in the ``sky'' from poor
background cancellation.  Aperture photometry was performed on GD 362 and
FS 147.  Results are shown in Table 1 where we use the flux calibration of magnitudes given by 
Tokunaga (2000).

 We compute the expected photospheric flux from the star, $F_{*}$, from the simple black-body expression:
\begin{equation}
F_{*}\;=\;{\pi}\left(\frac{R_{*}}{D_{*}}\right)^{2}\,B_{\nu}(T_{eff})
\end{equation}
According to Bergeron et al. (1995) and Rohrmann (2001), the atmospheres of hydrogen-rich white dwarfs
in the temperature and wavelength range of interest here can be well reproduced
by simple black bodies.  Gianninas et al. (2004) estimate that
the distance to GD 362 is somewhere between 22 and 26 pc. We adopt $D_{*}$ = 25 pc in order to best reproduce the $J$ band data by purely photospheric emission for a star with $T_{eff}$ = 9740 and $R_{*}$ = 3.5 ${\times}$ 10$^{8}$ cm, derived  from the mass and gravity given by Gianninas et al. (2004).  The total fluxes with the photospheric and excess contributions are listed in Table 1 and displayed  in Figure 2.

Although our wavelength coverage is incomplete, we find that the observed integrated flux of the infrared excess is ${\sim}$3\% of the total flux received from the star.
This result substantially constrains any model for the dust emission.

\section{MODEL FOR THE DUST DISTRIBUTION AND EMISSION}

We now consider models to account for the infrared excess around GD 362.
From the radius and effective temperature given above, the
total luminosity of GD 362 is 2.0 ${\times}$ 10$^{-4}$ L$_{\odot}$.  Thus the
process producing the infrared excess has a luminosity of 6 ${\times}$ 10$^{-6}$ L$_{\odot}$.  If the excess is emitted by a companion, this secondary object
must be substellar.    

 The evolutionary lifetime to the tip of the Asymptotic Giant Branch of the ${\sim}$ 7 M$_{\odot}$ main-sequence progenitor of a 1.2 M$_{\odot}$ white dwarf (Weidemann 2000) is likely to be less than 0.1 Gyr  (Girardi et al. 2000), 
 so that the white dwarf cooling age is an excellent proxy for its total age. According to Garcia-Berro et al. (1997), the cooling time
of this 1.2 M$_{\odot}$ white dwarf is about 5  Gyr.  However, the cooling age
of a white dwarf with a  luminosity of 2.0 ${\times}$ 10$^{-4}$ L$_{\odot}$ can be as low as ${\sim}$ 2 Gyr, depending upon the star's mass (Salaris et al. 1997, Hansen 2004).    
Burrows et al. (1993) show that some  brown dwarfs  with ages between 2 and 5 Gyr   have luminosities near 10$^{-5}$ L$_{\odot}$.  Their models also show that such brown dwarfs
have radii near 6 ${\times}$ 10$^{9}$ cm.  
The infrared excess around GD 362 can be roughly reproduced by
a black body with a color temperature of 600 K.  Such an object would require a radius of ${\sim}$2 ${\times}$ 10$^{10}$ cm, much larger than expected for
cooling brown dwarfs.   Thus, it is very unlikely that the infrared excess around
GD 362 is explained by direct emission from either a stellar or substellar companion.

A plausible model for the star's infrared
excess is that  circumstellar dust reprocesses
3\% of the emitted starlight.  To account for this result, we  follow Jura (2003) and adopt a simple model  of  a passive, flat, opaque dust
disk orbiting GD 362.   The white dwarf's gravity is so strong that a standard estimate of the disk thickness of
1200 K gas orbiting at 10$^{10}$ cm from the star yields a height much less than the radius of the white dwarf.  Therefore, the disk is taken as geometrically flat.   As a first approximation, the dust at distance $R$ from the star is  characterized by a single temperature, $T$, given by the expression (see, for example, Chiang \& Goldreich 1997):
\begin{equation}
T\;{\approx}\;\left(\frac{2}{3{\pi}}\right)^{1/4}\,\left(\frac{R_{*}}{R}\right)^{3/4}\;T_{eff}
\end{equation}
With this temperature profile, the predicted flux, $F_{d}$, from the disk at the Earth,  is (Jura 2003):
\begin{equation}
F_{d}\;{\approx}\;12\,{\pi}^{1/3}\,\left(\frac{R_{*}^{2}\,cos\,i}{D_{*}^{2}}\right)\,\left(\frac{2\,k_{B}T_{eff}}{3\,h{\nu}}\right)^{8/3}\left(\frac{h{\nu}^{3}}{c^{2}}\right)\,{\int}_{x_{in}}^{x_{out}}\frac{x^{5/3}}{e^{x}\;-\;1}\,dx
\end{equation}
   In equation (3), we define 
$x_{in}$ = $h{\nu}/kT_{in}$ and $x_{out}$ = ${\infty}$, corresponding to an outer temperature near 0 K or, equivalently, an infinitely extended disk.  Our model is insensitive to this outer boundary temperature because  cool dust does not emit appreciably
in the infrared bands for which we have data. With longer wavelength observations,
we could constrain the outer structure of the circumstellar dust.
We adopt an inner dust temperature of 1200 K, approximately where refractory silicate dust rapidly sublimates.  We find that $T$ =  1200 K occurs at $R$ = 9.7 $R_{*}$ or 3.4 ${\times}$ 10$^{9}$ cm. We assume that the disk is face-on so that $\cos\,i$ = 1. 

We display in Figure 2 the comparison between the data for the infrared excess
and the model.  Given the simplicity of the model and that the only unconstrained parameter is the inclination of the disk, the approximate agreement
 with the data is sufficiently close that our model is viable.  However, the model fails somewhat at $L'$ and by about a factor of 2 to account
for the observed $N'$ flux.  One possibility is that the dust
around GD 362 has a prominent silicate emission feature as has been found for G29-38 (M. Kuchner and T. von Hippel, private communication) and even possibly  PAH emission features.  Emission lines are produced by an opaque disk if the  warmest dust lies on the surface of the disk as can occur for a system heated from
above. Another possibility is that the accretion rate onto the white dwarf is
sufficiently high that dissipative heating of the disk is important and
that the star's luminosity is also enhanced (see below).  These effects would lead to more emission from the disk than predicted by our model.    

\section{DISK PARAMETERS}
Above, we have shown that the infrared excess around GD 362 can be modeled as a flat, opaque disk. Here, we consider how the disk might have formed.

 The amount of mass in this disk need not be very large compared, say, to the mass of a single massive Solar System comet.    If the
disk extends to ${\sim}$ 10$^{10}$ cm, then its total area may be ${\sim}$ 3 ${\times}$ 10$^{20}$ cm$^{2}$.  In order for the disk to be opaque even at 10 ${\mu}$m, the mass column of dust needs to be at least 10$^{-3}$ g cm$^{-2}$ (Ossenkopf et al. 1992).  Thus the  mass of the disk is greater or equal to ${\sim}$ 3 ${\times}$ 10$^{17}$ g.  However, since the disk is  opaque, we can only place a lower
bound on its mass; it could be orders of magnitude greater than this minimum. 

It is likely that dust in the disk around GD 362 accretes onto the star and accounts for the metals in its photosphere. The required rate of accretion, ${\dot M}_{ac}$ is uncertain because suitable models have not been published for
white dwarfs as massive as 1.2 M$_{\odot}$.       Extrapolation of the published models by Paquette et al. (1986) for  stars with $T_{eff}$ = 10000 K and 0.2 M$_{\odot}$, 0.6 M$_{\odot}$ and 1.0 M$_{\odot}$  suggests
 that in order to produce the observed essentially solar calcium abundance, then   ${\dot M_{ac}}$ onto GD 362 may be ${\sim}$10$^{11}$ g s$^{-1}$. 

The inferred accretion rate onto GD 362 is sufficiently high that it could appreciably contribute to
the star's luminosity.
The total accretion luminosity, $L_{ac}$, is given by 
\begin{equation}
L_{ac}\;=\;\frac{G\,M_{*}{\dot M_{ac}}}{R_{*}}   
\end{equation}
Thus, if GD 362 only accretes metals from
grains, then $L_{ac}$ = 1 ${\times}$ 10$^{-5}$ L$_{\odot}$, about 5\% of the star's bolometric luminosity.  If, however, the star also accretes gas with 100 times the mass of the heavy metals, then the total luminosity from accretion would exceed the bolometric luminosity of the star by a factor of 5.   It seems unlikely that
GD 362 is accreting from the gas-rich interstellar medium.

Following the discussion in Davidsson (1999), we note that any asteroid
that passed  ${\leq}$ 10$^{11}$ cm from GD 362 would venture within the Roche radius and be tidally disrupted.
The debris from this event could eventually form a disk, an event analogous to scenarios for the formation of
the rings around Saturn and other planets in our Solar System (see, for example, Dones 1991).  Debes \& Sigurdsson (2002) have described how the orbits of planets and asteroids are rearranged and become unstable when a star loses mass and becomes a white dwarf.  It is possible that this picture may explain why
an asteroid or planet could venture so near GD 362.  However, with a cooling age of 5 Gyr, by now, the system should  have achieved dynamic stability and the probability may be very low that an asteroid or
planet recently had its orbit dramatically altered.
Since GD 362 is accreting ${\sim}$ 10$^{11}$ g s$^{-1}$ and it has a cooling age
of 5 ${\times}$ 10$^{9}$ yr, then either the hypothetical disrupted parent body
had a mass of ${\sim}$ 2 $M_{\oplus}$ or the current disk has a lifetime appreciably less than the cooling age of the white dwarf.  

\section{CONCLUSIONS}
 GD 362 is a second white dwarf to be found to have an infrared excess that amounts to about
3\% of its bolometric luminosity.
This excess emission can be approximately explained as an opaque disk of refractory dust with an inner temperature of 1200 K.    
The circumstellar disk lies within the Roche radius of the white dwarf; the
dust might arise from the tidal disruption of some larger parent body. 
Accretion from the dust disk probably accounts for the star's very large photospheric abundance of calcium.          

Telescope time used for this
project was awarded by both the University of California (PI Becklin) and NASA
Keck (PI Weinberger) time allocation committees and traded for time on
Gemini. The $L'$-band data were taken with Gemini Director's Discretionary
time. Our results are in part based on observations obtained at the Gemini Observatory,
which is operated by the Association of Universities for
Research in Astronomy, Inc., under a cooperative agreement
with the NSF on behalf of the Gemini partnership: the National
Science Foundation (United States), the Particle Physics and
Astronomy Research Council (United Kingdom), the National
Research Council (Canada), CONICYT (Chile), the Australian
Research Council (Australia), CNPq (Brazil) and CONICET
(Argentina).
 We thank Scott Fisher and Tom Geballe for their help
obtaining the MICHELLE data.  This work has been partly supported by NASA.

\newpage
\begin{center}
TABLE 1 -- INFRARED DATA 
\end{center}
\begin{center}
\begin{tabular}{llllll}
\hline
\hline
${\lambda}$ & $m$ & $F_{obs}$ & $F_{*}$ & $F_{ex}$  \\
(${\mu}$m) &  (mag) & (mJy)   \\
\hline
1.22$^{a}$ ($J$) & 16.09 ${\pm}$ 0.03 & 0.60 ${\pm}$ 0.020& 0.60 & 0.00\\
1.22$^{b}$ ($J$) & 16.16 ${\pm}$ 0.09 \\
1.65$^{a}$ ($H$) & 15.99 ${\pm}$ 0.03 & 0.42 ${\pm}$ 0.01 & 0.39 & 0.03\\
1.65$^{c}$ ($H$) & 15.95 ${\pm}$ 0.06 \\
2.16$^{c}$ ($K_{s}$) & 15.85 ${\pm}$ 0.06 \\
2.18$^{a}$ ($K$) & 15.86 ${\pm}$ 0.03 & 0.30 ${\pm}$ 0.01 & 0.26 & 0.04\\
3.76$^{a}$ ($L'$) & 14.2 ${\pm}$ 0.3 & 0.52 ${\pm}$ 0.13 & 0.10 & 0.42\\
11.3$^{a}$ ($N'$) &     & 1.4 ${\pm}$ 0.3 & 0.01 & 1.4\\
\hline
\end{tabular}
\end{center}
$F_{obs}$ is the observed flux plotted as points in Figure 2.  
$^{a}$measured at Gemini North telescope, $^{b}$taken from the 2MASS catalog, $^{c}$measured at Magellan I telescope.  Since they are only upper limits, we do not list the 2MASS fluxes for $H$ and $K$.
\newpage
\begin{figure}
\epsscale{1}
\plotone{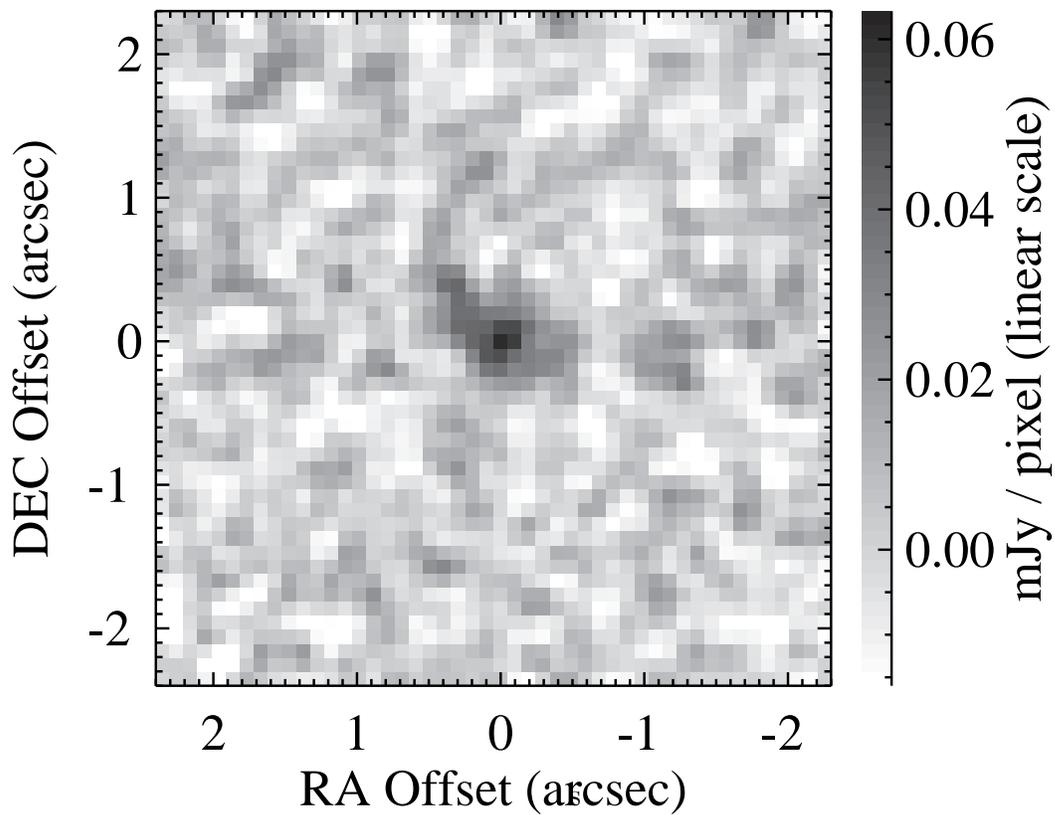}
\vspace{20pts}
\caption{The unresolved MICHELLE $N'$ image of  GD 362 which was smoothed by convolution with a Gaussian for display purposes.  The small amount of flux to the NE of GD 362 is consistent with the level of noise in the image and does not significantly affect the aperture photometry.}
\end{figure}
\begin{figure}
\plotone{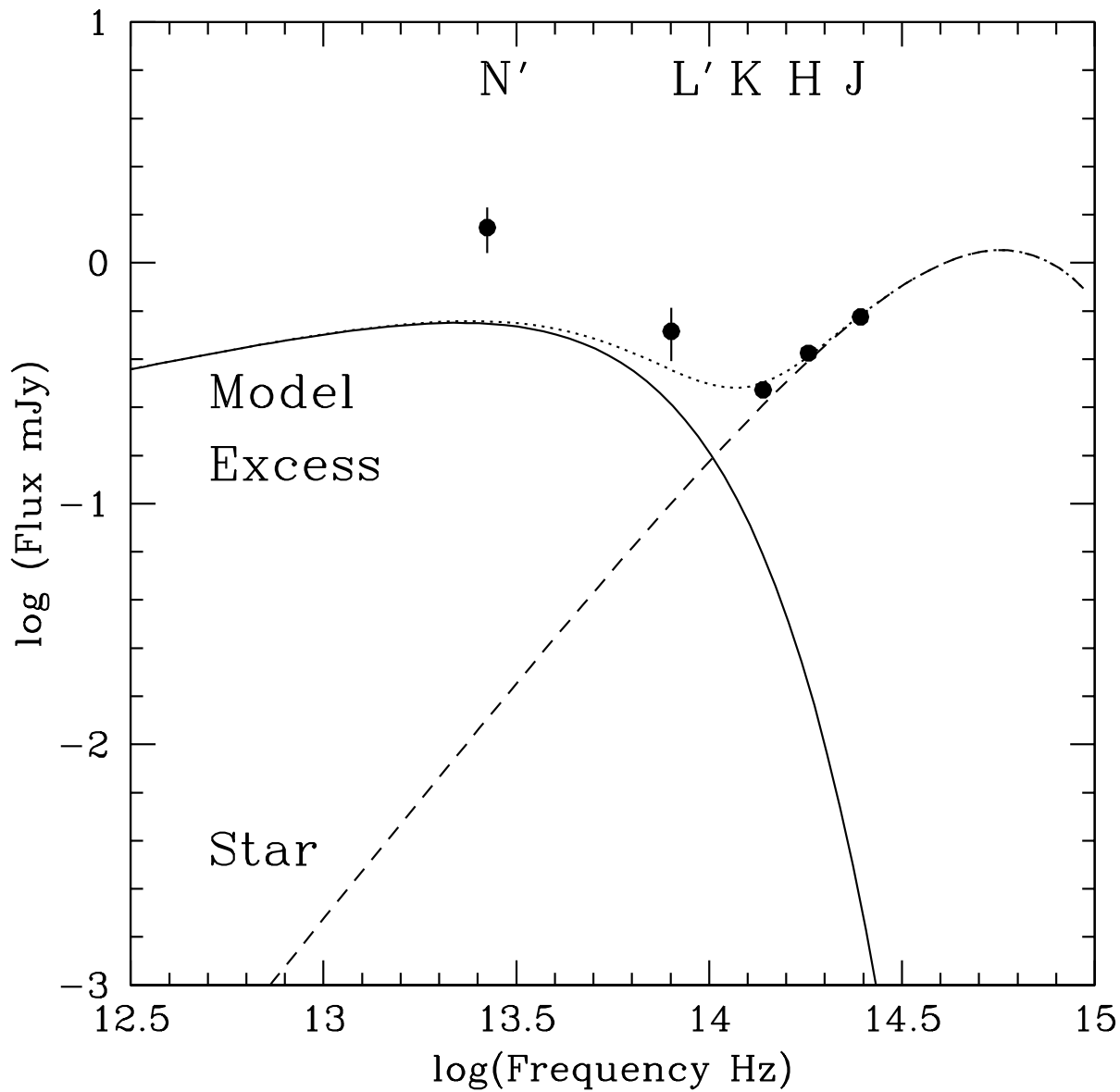}
\caption {Data from Table 1 are  compared with our model spectral energy distribution for the circumstellar  dust around GD 362.
   The solid  line shows the 
model calculation from equation (3)  where the outermost grains have a temperature of 0 K and  the  inner disk temperature is 1200 K (see text).  The long-dashed line shows the emission from the star while the short-dashed line shows the sum
of the stellar and the modeled circumstellar fluxes.}
\end{figure}
\end{document}